\documentclass[aps,pre,showpacs,twocolumn,superscriptaddress,letter]{revtex4}
\usepackage{graphics} 
\usepackage{graphicx}
\usepackage{epsfig}
\usepackage{amsmath,amssymb,wasysym,epsfig,capt-of,ifthen,calc}
\usepackage{bbm} 
\usepackage{latexsym} %extra symbols
\usepackage{setspace} %doublespacing 
\usepackage{array} %for some equations 
\usepackage{delarray} %for some equations
\usepackage{afterpage} \usepackage{graphicx}% Include figure files
\usepackage{dcolumn}% Align table columns on decimal point
\usepackage{bm}% bold math \usepackage{array} \usepackage{hyperref}
\newcommand{\be}{\begin{equation}}
\newcommand{\ee}{\end{equation}}
\newcommand{\bea}{\begin{eqnarray}}
\newcommand{\eea}{\end{eqnarray}}

\begin{document}

% Use the \preprint command to place your local institutional report
% number in the upper righthand corner of the title page in preprint mode.
% Multiple \preprint commands are allowed.
% Use the 'preprintnumbers' class option to override journal defaults
% to display numbers if necessary
%\preprint{}

%Title of paper
\title{Clustering Drives Assortativity and Community Structure in Ensembles of Networks}

\author{David V. Foster} \email[]{ventres@gmail.com}
\affiliation{Complexity Science Group, University of Calgary, Calgary
 T2N 1N4, Canada }

\author{Jacob G. Foster} 
\affiliation{Department of Sociology, University of Chicago, Chicago 60615, USA }
 
 \author{Peter Grassberger} 
\affiliation{Complexity Science Group, University of Calgary, Calgary
 T2N 1N4, Canada }
 \affiliation{NIC, Forschungszentrum J\"ulich, D-52425 J\"ulich, Germany}

\author{Maya Paczuski} 
\affiliation{Complexity Science Group, University of Calgary, Calgary
 T2N 1N4, Canada }

% repeat the \author .. \affiliation  etc. as needed
% \email, \thanks, \homepage, \altaffiliation all apply to the current
% author. Explanatory text should go in the []'s, actual e-mail
% address or url should go in the {}'s for \email and \homepage.
% Please use the appropriate macro foreach each type of information

% \affiliation command applies to all authors since the last
% \affiliation command. The \affiliation command should follow the
% other information
% \affiliation can be followed by \email, \homepage, \thanks as well.

%Collaboration name if desired (requires use of superscriptaddress
%option in \documentclass). \noaffiliation is required (may also be
%used with the \author command).
%\collaboration can be followed by \email, \homepage, \thanks as well.
%\collaboration{}
%\noaffiliation

\date{\today}

\begin{abstract}
Clustering, assortativity, and communities are key features of complex networks. We probe dependencies between these attributes and find that ensembles with strong  clustering display both high assortativity by degree and prominent community structure, while  ensembles with high assortativity are much less biased towards clustering or community structure. Further, clustered networks can amplify small homophilic bias for trait assortativity. This marked asymmetry suggests that transitivity, rather than homophily, drives the standard nonsocial/social network dichotomy.
\end{abstract}

% insert suggested PACS numbers in braces on next line
% insert suggested keywords - APS authors don't need to do this
%\keywords{cell differentiation, genetic regulatory networks, bifurcations}

%\maketitle must follow title, authors, abstract, \pacs, and \keywords
\maketitle

% body of paper here - Use proper section commands
% References should be done using the~\cite, \ref, and \label commands
Networks provide convenient representations for diverse phenomena spanning physical, technological, social, biological and informational domains~\cite{ broder::2000, boccaletti, barabasi2004network, newman::2003C}.  They are often  complicated, historically contingent assemblies created by nonlinear processes.  Just as it is meaningful to ``explain" features of real networks with simple generative mechanisms, it is also informative to ask what properties to expect given no other information about a network save that it has a certain set of properties. 

In fact, network properties can be markedly interdependent~\cite{soffer::2005, holme::2007}.   We focus on three key features of undirected networks:  (1) the clustering coefficient, $C$, which reflects the tendency of the network to form triangles (transitivity)~\cite{ watts::1998, newman::2003B}; (2) the assortativity, $r$, which reflects the tendency of similar nodes to connect to one another (homophily)~\cite{newman::2002}; and (3) the modularity, $Q$, which reflects the tendency of nodes to form tightly interconnected communities~\cite{newman::2004B}.  

We show that ensembles of networks constrained by a transitive bias to be strongly clustered also become highly degree-assortative and modular. In contrast, ensembles constrained by a homophilic bias to be highly assortative show only weak clustering or modularity.  Hence, at the ensemble level a fundamental asymmetry exists between transitivity and homophily.  This asymmetry holds unless the distribution of the number of links attached to each node (the node's degree) is extremely broad.  Furthermore, a transitive bias can amplify the effect of a homophilic bias towards trait (i.e. race, age, education, etc.) assortativity \cite{kossinets2009origins} in network ensembles.

High values for the clustering, assortativity, and modularity are often observed in real-world social networks, while nonsocial networks may have low values ~\cite{newman::2003D}.  Although extensive social science literature posits homophily to be a dominant force in social network formation~\cite{mcpherson::2001, kossinets2009origins} (since social networks are highly assortative), our results show that a bias for transitive relationships (also called ``triadic closure" in sociology literature \cite{rapoport::1953}) is sufficient to obtain this effect in network ensembles. Our work is complementary to that of Newman and Park who produce assortativity and clustering characteristic of social networks by introducing modularity \cite{newman::2003D}.

\begin{figure}[!h]
\vskip -.25 cm
\includegraphics[scale=.24]{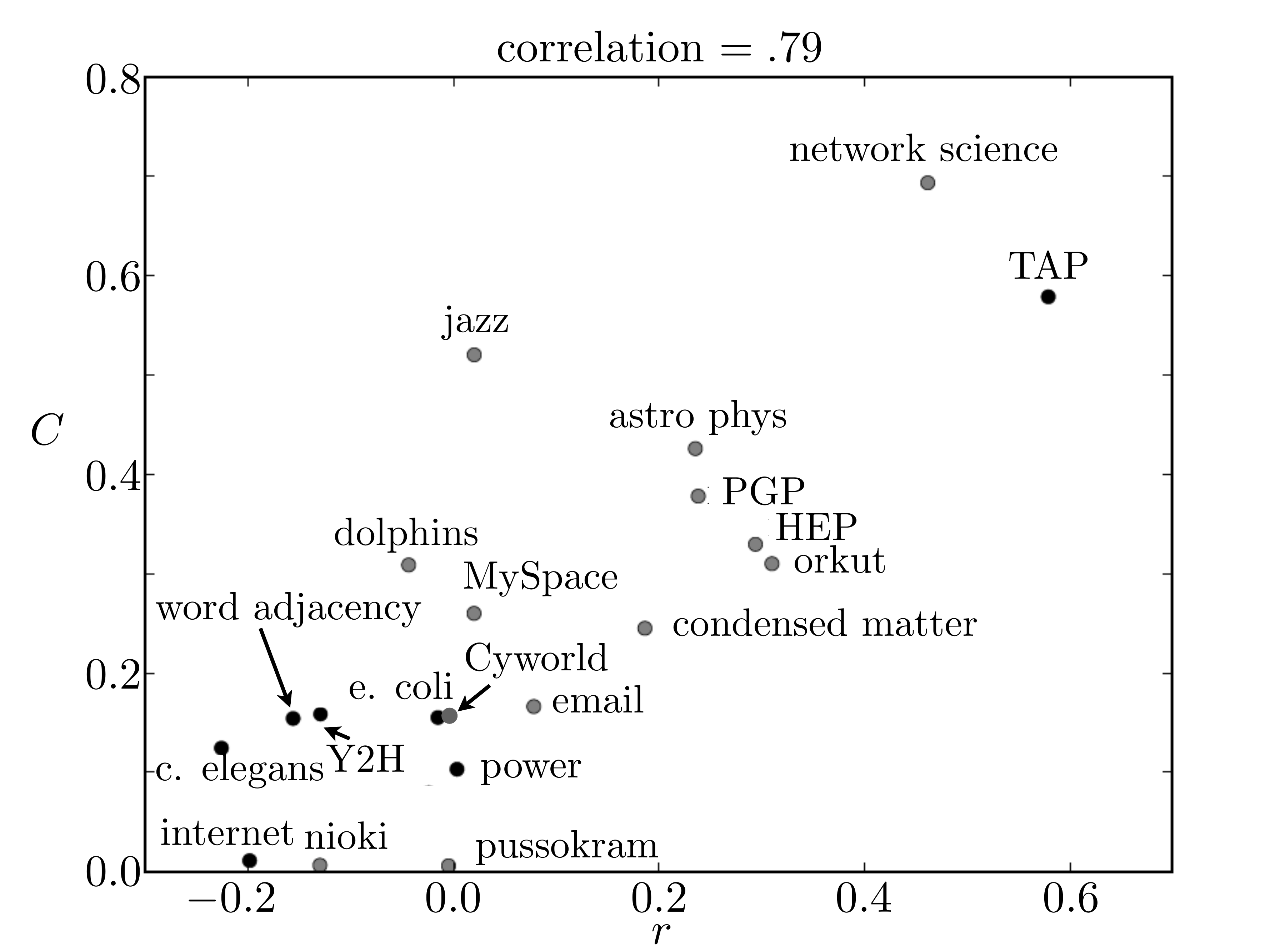}
\caption{\label{EmpiricalRvC} The relationship between the clustering coefficient, $C$, and the assortativity, $r$. Gray points represent social networks, black points represent other types of networks. {\bf Social networks}: astro phys (scientific collaboration)~\cite{newman::2001C} ; condensed matter (scientific collaboration)~\cite{newman::2001C}; Cyworld (online social)~\cite{ahn::2007}; dolphins (friendship)~\cite{lusseau::2003}; email (communication)~\cite{guimer::2003}; HEP (scientific collaboration)~\cite{newman::2001C}; jazz (musical collaboration)~\cite{gleiser::2003}; MySpace (online social)~\cite{ahn::2007};  network science (scientific collaboration)~\cite{newman::2006A}; nioki (online social)~\cite{holme::2004}; orkut (online social)~\cite{ahn::2007}; PGP (communication network)~\cite{bogu::2004}; pussokram (online dating)~\cite{holme::2004}. {\bf Non-social networks}: c. elegans (neural)~\cite{duch::2005}; e. coli (metabolic)~\cite{jeong::2000}; internet (router level)~\cite{newman_network_online}; power (connections between power stations)~\cite{watts::1998}; TAP (yeast protein-protein binding)~\cite{gavin::2002}; word adjacency (in English text)~\cite{newman::2006A}; Y2H (yeast protein-protein binding)~\cite{jeong::2001}.}
\vskip -.4 cm
\end{figure}

To begin, we note a distinct empirical correlation between $C$ and $r$ in real networks  illustrated in Fig.~\ref{EmpiricalRvC}, with social networks (generally) in the high $C$, high $r$ corner, and non-social networks (generally) in the low $C$, low $r$ one. The pattern suggests an  interdependence between the two features that transcends a simple nonsocial/social dichotomy.   For instance, consider  two networks in Fig.~\ref{EmpiricalRvC}: TAP is a high $C$, high $r$ protein-protein interaction network, generated by tandem affinity purification experiments~\cite{puig::2001}; Y2H is a \textit{weakly} clustered, \textit{disassortative} protein-protein interaction network, generated using yeast two hybridization~\cite{fields::1989}. The experimental methodology, by itself, can explain the difference, since TAP pulls out bound complexes and assigns links to every pair of proteins in the complex while Y2H tests each pair of proteins individually for direct binding.  Since transitivity has a natural origin in the construction of the TAP network, it is likely that the observed assortativity arises solely as a byproduct of the interrelationship between transitivity and assortativity rather than any direct homophilic tendency between proteins.

Since network properties often depend conspicuously on the degree sequence -- or the number of links attached to each node~\cite{newman::2001A} --  we consider ensembles of networks constrained to have the same fixed degree sequence (FDS).  Three real world networks are studied in detail: a collaboration network of high energy physicists (HEP)~\cite{newman::2001C}; a collaboration network of network scientists (NetSci)~\cite{newman::2006A}; and an encrypted communication network (PGP)~\cite{bogu::2004}.  We also examine  a randomly generated Erd\H{o}s - R\'enyi network (ER)~\cite{erdos::1959}.  Basic network parameters are given in Table \ref{Table1}. 

\begin{table}
\caption{\label{Table1} Important values for the empirical networks}
\begin{tabular} { p{1cm}  p{1cm}  p{1cm}  p{1.3cm}  p{1cm}  p{1cm}  p{.7cm} }
Name & \textit{N} & \textit{L} & \textit{r} & \textit{C} & \textit{Q} & Ref \\ \hline \hline
ER & 19680 & 41000 & -1.3e-5 & .00021 & .246 &~\cite{erdos::1959} \\ \hline
HEP & 7610 & 15751 & .29 & .33 & .40 &~\cite{newman::2001C} \\ \hline
NetSci & 1461 & 2742 & .46 & .70 & .47 &~\cite{newman::2006A} \\ \hline
PGP & 10680 & 24316 & .24 & .38 & .41 & ~\cite{bogu::2004} \\ \hline
\end{tabular}
\vskip -.3cm
\end{table}

We use a rewiring procedure~\cite{maslov::2002,foster::2007} to sample from each ensemble.  At each step of the procedure two links are chosen at random and their endpoints are exchanged, unless this would create a double link, in which case the step is skipped. This move set preserves the degree of each node but otherwise randomizes  connections.  To sample ensembles with specific features, we use a network Hamiltonian $H(G)$~\cite{milo::2002, lassig::2002, park::2004, foster::2010A}  to define an exponential ensemble by assigning a sampling weight $P(G) \propto e^{-H(G)}$ to each graph $G$.  Here we consider ensembles where $H(G)$ depends on $C$, $r$ and/or trait assortativity defined below.  Denoting the number of triangles in $G$ by $n_\Delta$, the degree of node $i$ by $k_i$, and the number of nodes by $N$, the clustering coefficient is defined as
\be 
C = {3n_\Delta\over {1\over 2}\sum_{i=1}^N (k_i-1)k_i}\quad .  \label{clust} 
\ee
Assortativity by degree is defined as the Pearson correlation coefficient between the degrees of nodes joined by a link~\cite{newman::2002}:
\be
   r = \frac{L\sum_{i=1}^L j_i k_i - [\sum_{i=1}^L j_i]^2}
              {L\sum_{i=1}^L j_i^2 - [\sum_{i=1}^L j_i]^2} \quad ,
\label{assort}
\ee
where $L$ is the number of links in the network and $j_i$ and $k_i$ are the degrees of nodes at each end of link $i$.

To get ensembles with specific values of $C$ or $r$  we use the following Hamiltonians:
\be
   H_{C'} = \beta |C' - C_{t}|, \:\: H_{r'} = \beta |r' - r_{t}|     \quad ,
\ee
where $C'$ is the current clustering coefficient  and $C_t$ is the target value, and similarly for $r'$. The parameter $\beta$ controls the strength of bias towards the target.  It is a transitive bias in $H_{C'}$ and a homophilic bias in $H_{r'}$.  

We employ simulated annealing based on a standard Metropolis-Hastings procedure with a rewiring move set~\cite{hastings::1970, barkema::1999}.  One pair of links in the network $G$ is switched to produce a new candidate network $G'$.  A valid move is accepted with probability
\be
   p = e^{H(G) - H(G')}    \quad         p\leq 1 \quad ,                         \label{prob}
\ee
and rejected with probability $1-p$. If $p>1$ the move is  accepted.  
Initially, the network is rewired $2\times 10^5$ times at $\beta=0$ to randomize links  and avoid strong hysteresis~\cite{foster::2010A}.  Then $\beta$ is increased slowly, rewiring $5\times 10^4$ times after each increase until $C$ (or $r$) hits $C_t$ (or $r_t$). The first network with $C = C_t$ ($r = r_t$) is a single sample from the ensemble of networks with a fixed degree sequence and $C = C_t$ ($r = r_t$).  The whole process then repeats, starting with the $\beta = 0$ quench. 

We also study the influence of transitivity on trait assortativity, $r_d$, which measures the tendency for nodes to connect to others with the same discrete trait (e.g. race, gender, etc.)~\cite{newman::2002}.  For this we add a homophilic bias $\beta_d$ for links between nodes with the same trait.  Defining $r_d \propto \sum_{\delta} e_{\delta\delta}$, where $e_{\delta\delta}$ is the fraction of links in the network from a node of type $\delta$ to another node of type $\delta$, the Hamiltonian becomes
\be
   H_d = \beta |C - C_{t}| + \beta_d \sum_{\delta} e_{\delta\delta} \quad .
\ee
Choosing different values of $C_{t}$ and $\beta_d$ allows one to explore how transitivity impacts trait assortativity at the ensemble level. 

\begin{figure}
\includegraphics[scale=.24]{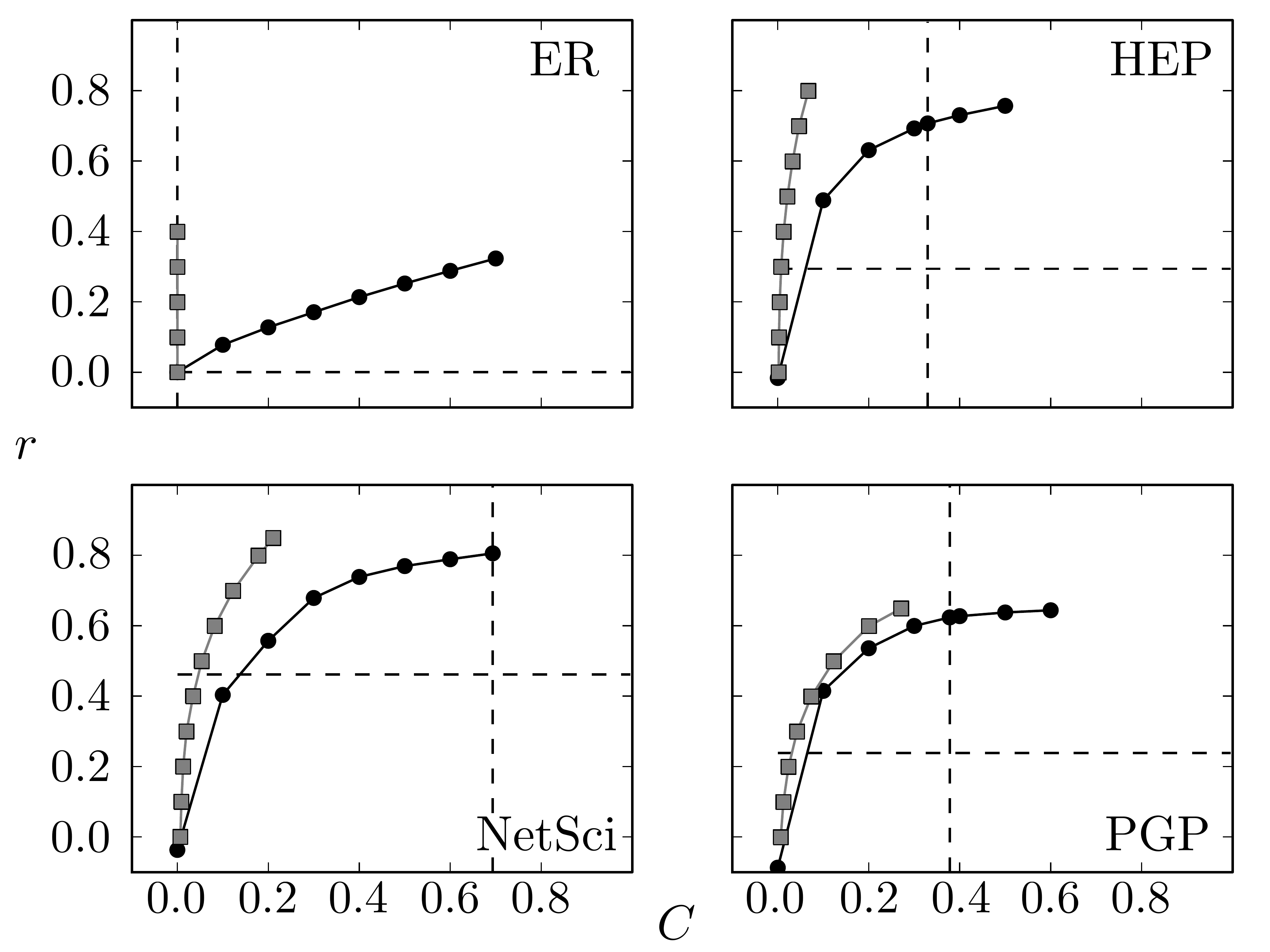}
\caption{\label{RvC} Controlling  assortativity (grey symbols) vs. controlling clustering coefficient (black symbols) for various network degree sequences.  $C$ is on the $x-$axis, $r$ on the $y-$axis.  Each point represents  average values from 100 samples  from an ensemble with specified $r$  or $C$  values. The dashed lines show the values of $r$ and $C$ for the original network. Note the asymmetry between the effect of $C$ on $r$ compared to $r$ on $C$.} 
\vskip -.6cm
\end{figure}

We examine ensembles constrained to have a particular value of $r$ (resp. $C$) and measure the value for the other feature $C$ (resp. $r$) averaged over $100$ samples from the ensemble.  Results are shown in Fig.~\ref{RvC}. The grey (resp. black) symbols show the values for ensembles with constrained $r$ (resp. $C$). Increasing transitivity to increase $C$ has a strong influence on $r$ in all cases, whereas increasing homophily to increase $r$ has relatively little impact on $C$. The asymmetry is strongest for narrow degree distributions (e.g. the ER network), and becomes less pronounced, but still apparent, as the degree distribution broadens. %For very broad distributions like the World Wide Web (data not shown), there is almost no difference between the influence of $r$ on $C$ and vice versa. 

The asymmetric relationship between $r$ and $C$ can be understood as follows:  For nodes to participate in as many transitive relationships as possible, their neighbours must be of similar degree.  Hence increasing clustering also increases $r$.  Increasing $r$ leads to links between nodes of similar degree, but these relationships need not be transitive.  For narrow degree distributions, one could divide all nodes of degree $k$ into two groups and only permit links between the two groups. Assortativity would be maximum, in the absence of any clustering.

On the other hand, for broad degree distributions (like PGP) only a few nodes of  high degree exist, but they have a large effect on $r$.  Hence for large $r$, the highest degree nodes are under strong pressure to link, thus creating many transitive relationships. Many social networks do not have broad degree distributions. In such cases homophily has only a weak influence on $C$ at the ensemble level.

Fig.~\ref{RvC} also indicates the $C$ and $r$ values for the real-world networks (dashed lines).  Ensembles of networks constrained to have the same $C$ as the real network exhibit far greater $r$.  Hence, social networks are actually {\it disassortative} relative to the ensemble of networks with the same clustering coefficient and degree sequence~\cite{foster::2010B}.  Indeed, the most likely way to create  many triangles is to densely interconnect the higher degree nodes so triangles clump together (as discussed in Ref.~\cite{foster::2010A}). Real social networks seem to spread clustering more evenly across the network, thus lowering $r$. For example in scientific collaboration networks, supervisory relationships may decrease the assortativity by creating links between lower degree students and higher degree professors.

\begin{figure}
\includegraphics[scale=.24]{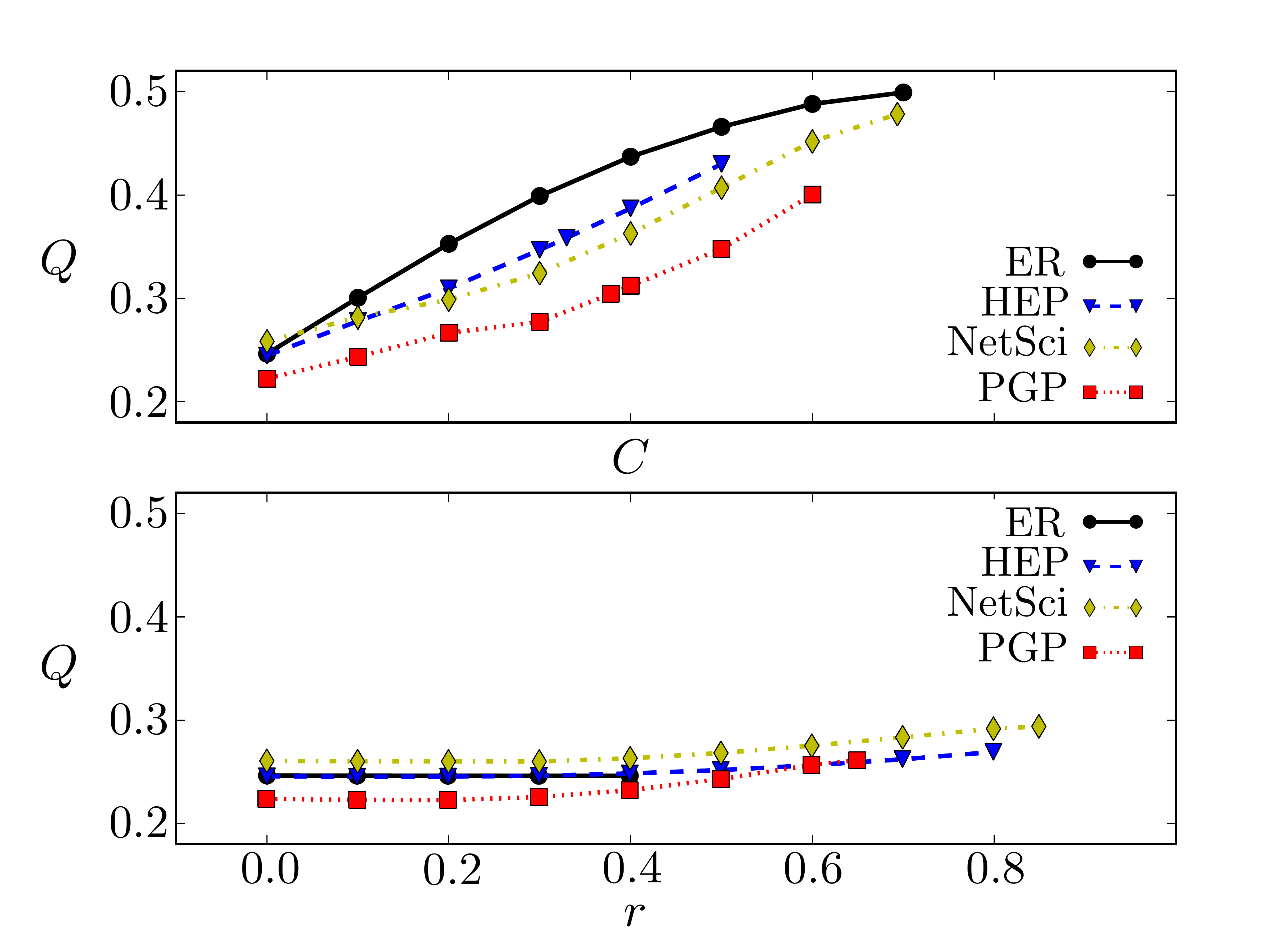}
\caption{\label{QvRC} (Color Online) Modularity $Q$ for various ensembles of networks with different target values for $C$ (top row) or  $r$ (bottom row). Clustering has a much larger impact on modularity than assortativity does.}
\vskip -.6cm
\end{figure}

We next consider the influence of $r$ and $C$ on modularity.  Many methods for extracting community structure exist~\cite{porter::2004, fortunato::2009}.  For definiteness, we use the one proposed by Newman and Girvan~\cite{newman::2004B}:  Given a partition of the network, $e_{ij}$ is the fraction of all edges connecting a node in community $i$ to one in community $j$, and $a_i = \sum_j e_{ij}$ is the fraction of all links within community $i$. The modularity of the network given partition $\mathcal{P}$ is defined as:
\be
   Q_{\mathcal{P}}= \sum_i (e_{ii} - a_i^2) \quad .
\label{mod}
\ee 
We use an agglomerative method~\cite{clauset::2004} to approximate the best partition and largest $Q_{\mathcal{P}}$, which we denote $Q$.

The top (resp. bottom) panel in Fig.~\ref{QvRC} shows the average $Q$ in ensembles with constrained $C$ (resp. $r$).  Transitivity has a more pronounced effect on modularity than does homophily.   The modularity achieved for the highly clustered ensembles approximates the actual modularity for the real networks (HEP, NetSci, and PGP; see Table~\ref{Table1}), unlike assortative ensembles without a transitive bias. 

\begin{figure}
\includegraphics[scale=.24]{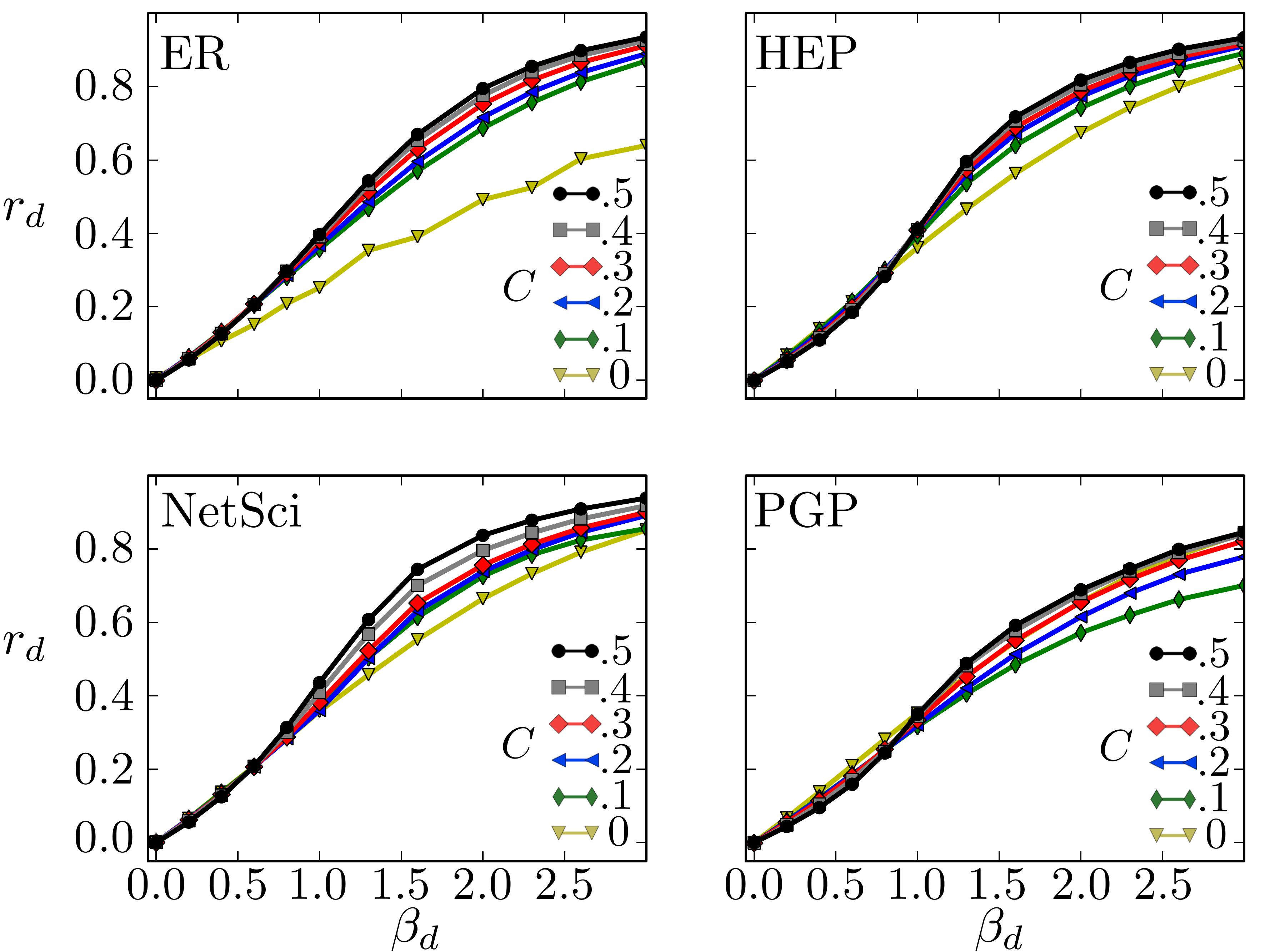}
\caption{\label{Trait} (Color Online) Trait assortativity $r_d$ (y-axis) for ensembles of networks with varying $C$ (indicated in the legend) and homophilic bias $\beta_d$ (x-axis).  For narrow degree distributions  clustering  amplifies the response of trait assortativity  on homophilic bias. For broad degree distribution the opposite occurs for small $\beta_d$.}
\vskip -.6cm
\end{figure}

Finally, we consider the effect of transitivity on trait assortativity, $r_d$. For each of the degree sequences, we create ensembles of networks with different target $C$ values and varying homophilic biases $\beta_d$. Since the actual data sets do not contain trait values, we assign each node one of three possible traits at random with equal probability. For ER, HEP, and NetSci we observe that ensembles with larger $C$ enhance $r_d$ relative to ensembles with the same homophilic bias but no clustering ($C=0$). This is especially clear for the narrowest (ER) degree sequence.  For the PGP network, which has a broad degree distribution, clustering appears to compete with the homophilic bias (e.g. the curves cross), leading to a more complicated scenario. The interdependence between clustering and trait assortativity thus appears to depend on the degree sequence, but for narrow degree sequences the positive relationship holds and transitivity enhances the effect of homophilic bias . We also note that increasing the trait assortativity of an ensemble had no impact on $C$, $r$, or $Q$ (data not shown).

We conjecture that the standard nonsocial/social (disassortative/assortative) dichotomy is driven by  transitive relationships in many social networks, such as in scientific collaborations.  As shown here, transitivity typically leads to assortativity.  This explains the anomalous position of TAP located within social networks, and is consistent with another anomaly in Fig.~\ref{EmpiricalRvC}: several online social networks show low clustering and low assortativity~\cite{hu::2009}. If assortative mixing by degree is the result of homophily by degree in social networks, this anomaly is hard to explain: why should popular people stop seeking each other out simply because the social network moved online? But if assortativity is a side-effect of transitivity, this effect is easier to understand: it is plausible that online social relationships are less transitive, since in the absence of spatially mediated interactions there is a smaller tendency to introduce mutual friends. We have not ruled out the scenario in reference \cite{newman::2003D}. Indeed, the causal factors driving network evolution are likely to be complex, multifaceted, and idiosyncratic. Our results on the asymmetric dependencies between clustering, assortativity, and modularity provide a warning about inferring causality from naive observations of network structure.

\bibliography{references.bib}

\begin{thebibliography}{44}
\expandafter\ifx\csname natexlab\endcsname\relax\def\natexlab#1{#1}\fi
\expandafter\ifx\csname bibnamefont\endcsname\relax
  \def\bibnamefont#1{#1}\fi
\expandafter\ifx\csname bibfnamefont\endcsname\relax
  \def\bibfnamefont#1{#1}\fi
\expandafter\ifx\csname citenamefont\endcsname\relax
  \def\citenamefont#1{#1}\fi
\expandafter\ifx\csname url\endcsname\relax
  \def\url#1{\texttt{#1}}\fi
\expandafter\ifx\csname urlprefix\endcsname\relax\def\urlprefix{URL }\fi
\providecommand{\bibinfo}[2]{#2}
\providecommand{\eprint}[2][]{\url{#2}}

\bibitem[{\citenamefont{Broder et~al.}(2000)}]{broder::2000}
\bibinfo{author}{\bibfnamefont{A.}~\bibnamefont{Broder}} \bibnamefont{et~al.},
  \bibinfo{journal}{Comp. Netw.} \textbf{\bibinfo{volume}{33}},
  \bibinfo{pages}{309} (\bibinfo{year}{2000}).

\bibitem[{\citenamefont{Boccaletti et~al.}(2006)}]{boccaletti}
\bibinfo{author}{\bibfnamefont{S.}~\bibnamefont{Boccaletti}}
  \bibnamefont{et~al.}, \bibinfo{journal}{Phys. Rep.}
  \textbf{\bibinfo{volume}{424}}, \bibinfo{pages}{175} (\bibinfo{year}{2006}).

\bibitem[{\citenamefont{Barab{\'a}si and Oltvai}(2004)}]{barabasi2004network}
\bibinfo{author}{\bibfnamefont{A.}~\bibnamefont{Barab{\'a}si}}
  \bibnamefont{and} \bibinfo{author}{\bibfnamefont{Z.}~\bibnamefont{Oltvai}},
  \bibinfo{journal}{Nat. Genet.} \textbf{\bibinfo{volume}{5}},
  \bibinfo{pages}{101} (\bibinfo{year}{2004}).

\bibitem[{\citenamefont{Newman}(2003{\natexlab{a}})}]{newman::2003C}
\bibinfo{author}{\bibfnamefont{M.~E.~J.} \bibnamefont{Newman}},
  \bibinfo{journal}{{SIAM} Review} \textbf{\bibinfo{volume}{45}},
  \bibinfo{pages}{167} (\bibinfo{year}{2003}{\natexlab{a}}).

\bibitem[{\citenamefont{Soffer and Vazquez}(2005)}]{soffer::2005}
\bibinfo{author}{\bibfnamefont{S.~N.} \bibnamefont{Soffer}} \bibnamefont{and}
  \bibinfo{author}{\bibfnamefont{A.}~\bibnamefont{Vazquez}},
  \bibinfo{journal}{Phys. Rev. E} \textbf{\bibinfo{volume}{71}},
  \bibinfo{pages}{057101} (\bibinfo{year}{2005}).

\bibitem[{\citenamefont{Holme and Zhao}(2007)}]{holme::2007}
\bibinfo{author}{\bibfnamefont{P.}~\bibnamefont{Holme}} \bibnamefont{and}
  \bibinfo{author}{\bibfnamefont{J.}~\bibnamefont{Zhao}},
  \bibinfo{journal}{Phys. Rev. E} \textbf{\bibinfo{volume}{75}},
  \bibinfo{pages}{046111} (\bibinfo{year}{2007}).

\bibitem[{\citenamefont{Watts and Strogatz}(1998)}]{watts::1998}
\bibinfo{author}{\bibfnamefont{D.~J.} \bibnamefont{Watts}} \bibnamefont{and}
  \bibinfo{author}{\bibfnamefont{S.~H.} \bibnamefont{Strogatz}},
  \bibinfo{journal}{Nature} \textbf{\bibinfo{volume}{393}},
  \bibinfo{pages}{440} (\bibinfo{year}{1998}).

\bibitem[{\citenamefont{Newman}(2003{\natexlab{b}})}]{newman::2003B}
\bibinfo{author}{\bibfnamefont{M.~E.~J.} \bibnamefont{Newman}},
  \bibinfo{journal}{Phys. Rev. E} \textbf{\bibinfo{volume}{68}},
  \bibinfo{pages}{026121} (\bibinfo{year}{2003}{\natexlab{b}}).

\bibitem[{\citenamefont{Newman}(2002)}]{newman::2002}
\bibinfo{author}{\bibfnamefont{M.~E.~J.} \bibnamefont{Newman}},
  \bibinfo{journal}{Phys. Rev. Lett.} \textbf{\bibinfo{volume}{89}},
  \bibinfo{pages}{208701} (\bibinfo{year}{2002}).

\bibitem[{\citenamefont{Newman and Girvan}(2004)}]{newman::2004B}
\bibinfo{author}{\bibfnamefont{M.~E.~J.} \bibnamefont{Newman}}
  \bibnamefont{and} \bibinfo{author}{\bibfnamefont{M.}~\bibnamefont{Girvan}},
  \bibinfo{journal}{Phys. Rev. E} \textbf{\bibinfo{volume}{69}},
  \bibinfo{pages}{026113} (\bibinfo{year}{2004}).

\bibitem[{\citenamefont{Kossinets and Watts}(2009)}]{kossinets2009origins}
\bibinfo{author}{\bibfnamefont{G.}~\bibnamefont{Kossinets}} \bibnamefont{and}
  \bibinfo{author}{\bibfnamefont{D.}~\bibnamefont{Watts}},
  \bibinfo{journal}{AJS} \textbf{\bibinfo{volume}{115}}, \bibinfo{pages}{405}
  (\bibinfo{year}{2009}).

\bibitem[{\citenamefont{Newman and Park}(2003)}]{newman::2003D}
\bibinfo{author}{\bibfnamefont{M.~E.~J.} \bibnamefont{Newman}}
  \bibnamefont{and} \bibinfo{author}{\bibfnamefont{J.}~\bibnamefont{Park}},
  \bibinfo{journal}{Phys. Rev. E} \textbf{\bibinfo{volume}{68}},
  \bibinfo{pages}{036122} (\bibinfo{year}{2003}).

\bibitem[{\citenamefont{McPherson et~al.}(2001)\citenamefont{McPherson,
  Smith-Lovin, and Cook}}]{mcpherson::2001}
\bibinfo{author}{\bibfnamefont{M.}~\bibnamefont{McPherson}},
  \bibinfo{author}{\bibfnamefont{L.}~\bibnamefont{Smith-Lovin}},
  \bibnamefont{and} \bibinfo{author}{\bibfnamefont{J.}~\bibnamefont{Cook}},
  \bibinfo{journal}{Annu. Rev. Sociol.} \textbf{\bibinfo{volume}{27}},
  \bibinfo{pages}{415} (\bibinfo{year}{2001}).

\bibitem[{\citenamefont{Rapoport}(1953)}]{rapoport::1953}
\bibinfo{author}{\bibfnamefont{A.}~\bibnamefont{Rapoport}},
  \bibinfo{journal}{Bull. Math. Biol.} \textbf{\bibinfo{volume}{15}},
  \bibinfo{pages}{523} (\bibinfo{year}{1953}).

\bibitem[{\citenamefont{Newman}(2001)}]{newman::2001C}
\bibinfo{author}{\bibfnamefont{M.~E.~J.} \bibnamefont{Newman}},
  \bibinfo{journal}{PNAS} \textbf{\bibinfo{volume}{98}}, \bibinfo{pages}{404}
  (\bibinfo{year}{2001}).

\bibitem[{\citenamefont{Ahn et~al.}(2007)\citenamefont{Ahn, Han, Kwak, Moon,
  and Jeong}}]{ahn::2007}
\bibinfo{author}{\bibfnamefont{Y.-Y.} \bibnamefont{Ahn}},
  \bibinfo{author}{\bibfnamefont{S.}~\bibnamefont{Han}},
  \bibinfo{author}{\bibfnamefont{H.}~\bibnamefont{Kwak}},
  \bibinfo{author}{\bibfnamefont{S.}~\bibnamefont{Moon}}, \bibnamefont{and}
  \bibinfo{author}{\bibfnamefont{H.}~\bibnamefont{Jeong}}, in
  \emph{\bibinfo{booktitle}{Proceedings of the 16th international conference on
  World Wide Web}} (\bibinfo{organization}{ACM}, \bibinfo{year}{2007}).

\bibitem[{\citenamefont{Lusseau et~al.}(2003)}]{lusseau::2003}
\bibinfo{author}{\bibfnamefont{D.}~\bibnamefont{Lusseau}} \bibnamefont{et~al.},
  \bibinfo{journal}{Behav. Ecol. Sociobiol.} \textbf{\bibinfo{volume}{54}},
  \bibinfo{pages}{396} (\bibinfo{year}{2003}).

\bibitem[{\citenamefont{Guimer\'{a} et~al.}(2003)\citenamefont{Guimer\'{a},
  Danon, {D�az-Guilera}, Giralt, and Arenas}}]{guimer::2003}
\bibinfo{author}{\bibfnamefont{R.}~\bibnamefont{Guimer\'{a}}},
  \bibinfo{author}{\bibfnamefont{L.}~\bibnamefont{Danon}},
  \bibinfo{author}{\bibfnamefont{A.}~\bibnamefont{{D�az-Guilera}}},
  \bibinfo{author}{\bibfnamefont{F.}~\bibnamefont{Giralt}}, \bibnamefont{and}
  \bibinfo{author}{\bibfnamefont{A.}~\bibnamefont{Arenas}},
  \bibinfo{journal}{Phys. Rev. E} \textbf{\bibinfo{volume}{68}},
  \bibinfo{pages}{065103} (\bibinfo{year}{2003}).

\bibitem[{\citenamefont{Gleiser and Danon}(2003)}]{gleiser::2003}
\bibinfo{author}{\bibfnamefont{P.~M.} \bibnamefont{Gleiser}} \bibnamefont{and}
  \bibinfo{author}{\bibfnamefont{L.}~\bibnamefont{Danon}},
  \bibinfo{journal}{Adv. Complex Sys.} \textbf{\bibinfo{volume}{6}},
  \bibinfo{pages}{565–573} (\bibinfo{year}{2003}).

\bibitem[{\citenamefont{Newman}(2006)}]{newman::2006A}
\bibinfo{author}{\bibfnamefont{M.~E.~J.} \bibnamefont{Newman}},
  \bibinfo{journal}{Phys. Rev. E} \textbf{\bibinfo{volume}{74}},
  \bibinfo{pages}{036104} (\bibinfo{year}{2006}).

\bibitem[{\citenamefont{Holme et~al.}(2004)\citenamefont{Holme, Edling, and
  Liljeros}}]{holme::2004}
\bibinfo{author}{\bibfnamefont{P.}~\bibnamefont{Holme}},
  \bibinfo{author}{\bibfnamefont{C.~R.} \bibnamefont{Edling}},
  \bibnamefont{and} \bibinfo{author}{\bibfnamefont{F.}~\bibnamefont{Liljeros}},
  \bibinfo{journal}{Soc. Networks} \textbf{\bibinfo{volume}{26}},
  \bibinfo{pages}{155} (\bibinfo{year}{2004}).

\bibitem[{\citenamefont{Bogu\~n\'a et~al.}(2004)\citenamefont{Bogu\~n\'a,
  Pastor-Satorras, D\'\i{}az-Guilera, and Arenas}}]{bogu::2004}
\bibinfo{author}{\bibfnamefont{M.}~\bibnamefont{Bogu\~n\'a}},
  \bibinfo{author}{\bibfnamefont{R.}~\bibnamefont{Pastor-Satorras}},
  \bibinfo{author}{\bibfnamefont{A.}~\bibnamefont{D\'\i{}az-Guilera}},
  \bibnamefont{and} \bibinfo{author}{\bibfnamefont{A.}~\bibnamefont{Arenas}},
  \bibinfo{journal}{Phys. Rev. E} \textbf{\bibinfo{volume}{70}},
  \bibinfo{pages}{056122} (\bibinfo{year}{2004}).

\bibitem[{\citenamefont{Duch and Arenas}(2005)}]{duch::2005}
\bibinfo{author}{\bibfnamefont{J.}~\bibnamefont{Duch}} \bibnamefont{and}
  \bibinfo{author}{\bibfnamefont{A.}~\bibnamefont{Arenas}},
  \bibinfo{journal}{Phys. Rev. E} \textbf{\bibinfo{volume}{72}},
  \bibinfo{pages}{027104} (\bibinfo{year}{2005}).

\bibitem[{\citenamefont{Jeong et~al.}(2000)\citenamefont{Jeong, Tombor, Albert,
  Oltvai, and Barab\'asi}}]{jeong::2000}
\bibinfo{author}{\bibfnamefont{H.}~\bibnamefont{Jeong}},
  \bibinfo{author}{\bibfnamefont{B.}~\bibnamefont{Tombor}},
  \bibinfo{author}{\bibfnamefont{R.}~\bibnamefont{Albert}},
  \bibinfo{author}{\bibfnamefont{Z.~N.} \bibnamefont{Oltvai}},
  \bibnamefont{and}
  \bibinfo{author}{\bibfnamefont{A.}~\bibnamefont{Barab\'asi}},
  \bibinfo{journal}{Nature} \textbf{\bibinfo{volume}{407}},
  \bibinfo{pages}{651} (\bibinfo{year}{2000}).

\bibitem[{\citenamefont{Newman}()}]{newman_network_online}
\bibinfo{author}{\bibfnamefont{M.~E.~J.} \bibnamefont{Newman}},
  \emph{\bibinfo{title}{Network data}},
  \bibinfo{howpublished}{http://www-personal.umich.edu/{\textasciitilde}mejn/n%
etdata/}.

\bibitem[{\citenamefont{Gavin et~al.}(2002)}]{gavin::2002}
\bibinfo{author}{\bibfnamefont{A.~C.} \bibnamefont{Gavin}}
  \bibnamefont{et~al.}, \bibinfo{journal}{Nature}
  \textbf{\bibinfo{volume}{415}}, \bibinfo{pages}{141} (\bibinfo{year}{2002}).

\bibitem[{\citenamefont{Jeong et~al.}(2001)\citenamefont{Jeong, Mason,
  Barab\'asi, and Oltvai}}]{jeong::2001}
\bibinfo{author}{\bibfnamefont{H.}~\bibnamefont{Jeong}},
  \bibinfo{author}{\bibfnamefont{S.~P.} \bibnamefont{Mason}},
  \bibinfo{author}{\bibfnamefont{A.}~\bibnamefont{Barab\'asi}},
  \bibnamefont{and} \bibinfo{author}{\bibfnamefont{Z.~N.}
  \bibnamefont{Oltvai}}, \bibinfo{journal}{Nature}
  \textbf{\bibinfo{volume}{411}}, \bibinfo{pages}{41} (\bibinfo{year}{2001}).

\bibitem[{\citenamefont{Puig et~al.}(2001)}]{puig::2001}
\bibinfo{author}{\bibfnamefont{O.}~\bibnamefont{Puig}} \bibnamefont{et~al.},
  \bibinfo{journal}{Methods} \textbf{\bibinfo{volume}{24}},
  \bibinfo{pages}{218} (\bibinfo{year}{2001}).

\bibitem[{\citenamefont{Fields and Song}(1989)}]{fields::1989}
\bibinfo{author}{\bibfnamefont{S.}~\bibnamefont{Fields}} \bibnamefont{and}
  \bibinfo{author}{\bibfnamefont{O.}~\bibnamefont{Song}},
  \bibinfo{journal}{Nature} \textbf{\bibinfo{volume}{340}},
  \bibinfo{pages}{245} (\bibinfo{year}{1989}).

\bibitem[{\citenamefont{Newman et~al.}(2001)\citenamefont{Newman, Strogatz, and
  Watts}}]{newman::2001A}
\bibinfo{author}{\bibfnamefont{M.~E.~J.} \bibnamefont{Newman}},
  \bibinfo{author}{\bibfnamefont{S.~H.} \bibnamefont{Strogatz}},
  \bibnamefont{and} \bibinfo{author}{\bibfnamefont{D.~J.} \bibnamefont{Watts}},
  \bibinfo{journal}{Phys. Rev. E} \textbf{\bibinfo{volume}{64}},
  \bibinfo{pages}{026118} (\bibinfo{year}{2001}).

\bibitem[{\citenamefont{Erd\H{o}s and R\'enyii}(1959)}]{erdos::1959}
\bibinfo{author}{\bibfnamefont{P.}~\bibnamefont{Erd\H{o}s}} \bibnamefont{and}
  \bibinfo{author}{\bibfnamefont{A.}~\bibnamefont{R\'enyii}},
  \bibinfo{journal}{Publ. Math. Debrecen} \textbf{\bibinfo{volume}{6}},
  \bibinfo{pages}{156} (\bibinfo{year}{1959}).

\bibitem[{\citenamefont{Maslov and Sneppen}(2002)}]{maslov::2002}
\bibinfo{author}{\bibfnamefont{S.}~\bibnamefont{Maslov}} \bibnamefont{and}
  \bibinfo{author}{\bibfnamefont{K.}~\bibnamefont{Sneppen}},
  \bibinfo{journal}{Science} \textbf{\bibinfo{volume}{296}},
  \bibinfo{pages}{910} (\bibinfo{year}{2002}).

\bibitem[{\citenamefont{Foster et~al.}(2007)\citenamefont{Foster, Foster,
  Grassberger, and Paczuski}}]{foster::2007}
\bibinfo{author}{\bibfnamefont{J.~G.} \bibnamefont{Foster}},
  \bibinfo{author}{\bibfnamefont{D.~V.} \bibnamefont{Foster}},
  \bibinfo{author}{\bibfnamefont{P.}~\bibnamefont{Grassberger}},
  \bibnamefont{and} \bibinfo{author}{\bibfnamefont{M.}~\bibnamefont{Paczuski}},
  \bibinfo{journal}{Phys. Rev. E} \textbf{\bibinfo{volume}{76}},
  \bibinfo{pages}{46112} (\bibinfo{year}{2007}).

\bibitem[{\citenamefont{Milo et~al.}(2002)\citenamefont{Milo, {Shen-Orr},
  Itzkovitz, Kashtan, Chklovskii, and Alon}}]{milo::2002}
\bibinfo{author}{\bibfnamefont{R.}~\bibnamefont{Milo}},
  \bibinfo{author}{\bibfnamefont{S.}~\bibnamefont{{Shen-Orr}}},
  \bibinfo{author}{\bibfnamefont{S.}~\bibnamefont{Itzkovitz}},
  \bibinfo{author}{\bibfnamefont{N.}~\bibnamefont{Kashtan}},
  \bibinfo{author}{\bibfnamefont{D.}~\bibnamefont{Chklovskii}},
  \bibnamefont{and} \bibinfo{author}{\bibfnamefont{U.}~\bibnamefont{Alon}},
  \bibinfo{journal}{Science} \textbf{\bibinfo{volume}{298}},
  \bibinfo{pages}{824} (\bibinfo{year}{2002}).

\bibitem[{\citenamefont{Berg and L{\"a}ssig}(2002)}]{lassig::2002}
\bibinfo{author}{\bibfnamefont{J.}~\bibnamefont{Berg}} \bibnamefont{and}
  \bibinfo{author}{\bibfnamefont{M.}~\bibnamefont{L{\"a}ssig}},
  \bibinfo{journal}{Phys. Rev. Lett.} \textbf{\bibinfo{volume}{89}},
  \bibinfo{pages}{228701} (\bibinfo{year}{2002}).

\bibitem[{\citenamefont{Park and Newman}(2004)}]{park::2004}
\bibinfo{author}{\bibfnamefont{J.}~\bibnamefont{Park}} \bibnamefont{and}
  \bibinfo{author}{\bibfnamefont{M.~E.~J.} \bibnamefont{Newman}},
  \bibinfo{journal}{Phys. Rev. E} \textbf{\bibinfo{volume}{70}},
  \bibinfo{pages}{066117} (\bibinfo{year}{2004}).

\bibitem[{\citenamefont{Foster et~al.}(2010{\natexlab{a}})\citenamefont{Foster,
  Foster, Paczuski, and Grassberger}}]{foster::2010A}
\bibinfo{author}{\bibfnamefont{D.~V.} \bibnamefont{Foster}},
  \bibinfo{author}{\bibfnamefont{J.~G.} \bibnamefont{Foster}},
  \bibinfo{author}{\bibfnamefont{M.}~\bibnamefont{Paczuski}}, \bibnamefont{and}
  \bibinfo{author}{\bibfnamefont{P.}~\bibnamefont{Grassberger}},
  \bibinfo{journal}{Phys. Rev. E} \textbf{\bibinfo{volume}{81}},
  \bibinfo{pages}{046115} (\bibinfo{year}{2010}{\natexlab{a}}).

\bibitem[{\citenamefont{{Hastings}}(1970)}]{hastings::1970}
\bibinfo{author}{\bibfnamefont{W.~K.} \bibnamefont{{Hastings}}},
  \bibinfo{journal}{Biometrika} \textbf{\bibinfo{volume}{57}},
  \bibinfo{pages}{97} (\bibinfo{year}{1970}).

\bibitem[{\citenamefont{Newman and Barkema}(1999)}]{barkema::1999}
\bibinfo{author}{\bibfnamefont{M.~E.~J.} \bibnamefont{Newman}}
  \bibnamefont{and} \bibinfo{author}{\bibfnamefont{G.~T.}
  \bibnamefont{Barkema}}, \emph{\bibinfo{title}{Monte Carlo methods in
  statistical physics}} (\bibinfo{publisher}{Oxford Univ. Press},
  \bibinfo{year}{1999}).

\bibitem[{\citenamefont{Foster et~al.}(2010{\natexlab{b}})\citenamefont{Foster,
  Foster, Paczuski, and Grassberger}}]{foster::2010B}
\bibinfo{author}{\bibfnamefont{J.~G.} \bibnamefont{Foster}},
  \bibinfo{author}{\bibfnamefont{D.~V.} \bibnamefont{Foster}},
  \bibinfo{author}{\bibfnamefont{M.}~\bibnamefont{Paczuski}}, \bibnamefont{and}
  \bibinfo{author}{\bibfnamefont{P.}~\bibnamefont{Grassberger}},
  \bibinfo{journal}{P. Natl. Acad. Sci. USA} \textbf{\bibinfo{volume}{107}},
  \bibinfo{pages}{10815} (\bibinfo{year}{2010}{\natexlab{b}}).

\bibitem[{\citenamefont{Porter}(2004)}]{porter::2004}
\bibinfo{author}{\bibfnamefont{C.}~\bibnamefont{Porter}}, \bibinfo{journal}{J.
  Computer-Mediated Comm.} \textbf{\bibinfo{volume}{10}}
  (\bibinfo{year}{2004}).

\bibitem[{\citenamefont{Fortunato}(2010)}]{fortunato::2009}
\bibinfo{author}{\bibfnamefont{S.}~\bibnamefont{Fortunato}},
  \bibinfo{journal}{Phys. Rep.} \textbf{\bibinfo{volume}{486}},
  \bibinfo{pages}{75} (\bibinfo{year}{2010}).

\bibitem[{\citenamefont{Clauset et~al.}(2004)\citenamefont{Clauset, Newman, and
  Moore}}]{clauset::2004}
\bibinfo{author}{\bibfnamefont{A.}~\bibnamefont{Clauset}},
  \bibinfo{author}{\bibfnamefont{M.~E.~J.} \bibnamefont{Newman}},
  \bibnamefont{and} \bibinfo{author}{\bibfnamefont{C.}~\bibnamefont{Moore}},
  \bibinfo{journal}{Phys. Rev. E} \textbf{\bibinfo{volume}{70}},
  \bibinfo{pages}{066111} (\bibinfo{year}{2004}).

\bibitem[{\citenamefont{Hu and Wang}(2009)}]{hu::2009}
\bibinfo{author}{\bibfnamefont{H.}~\bibnamefont{Hu}} \bibnamefont{and}
  \bibinfo{author}{\bibfnamefont{X.}~\bibnamefont{Wang}},
  \bibinfo{journal}{Europhys. Lett.} \textbf{\bibinfo{volume}{86}},
  \bibinfo{pages}{18003} (\bibinfo{year}{2009}).

\end{thebibliography}

\end{document}